\documentclass[prd,superscriptaddress,amsfonts,amssymb,amsmath,showpacs,twocolumn]{revtex4-2}
\usepackage{bm}
\usepackage{mathrsfs}
\usepackage{amsfonts}
\usepackage{latexsym}
\usepackage{graphicx}
\usepackage{amsmath}
\usepackage{palatino}
\usepackage{mathpazo}
\usepackage{textcomp}
\usepackage{float}
\usepackage{booktabs}
\usepackage{dcolumn}
\usepackage[table]{xcolor}
\usepackage{booktabs}
\usepackage{multirow}
\usepackage{hyperref}
\hypersetup{colorlinks,citecolor=blue}
\usepackage{amsmath}
\usepackage{xcolor}
\usepackage{orcidlink}
\usepackage{epsfig}
\usepackage{ragged2e}
\usepackage{caption}
\usepackage{subcaption}
\usepackage{commath}
\hypersetup{colorlinks,citecolor=blue}
\hypersetup{colorlinks=true,linkcolor=magenta,filecolor=magenta,    urlcolor=blue}

\def\be{\begin{equation}}
\def\ee{\end{equation}}
\def\bea{\begin{eqnarray}}
\def\eea{\end{eqnarray}}

\begin{document}
\title{Influence of external magnetic fields on charged particle motion around a Schwarzschild-like black hole}

\author{Sojida Mannobova}
\email{sojidamannabova2208@gmail.com}
\affiliation{National Research University TIIAME, Kori Niyoziy 39, Tashkent 100000, Uzbekistan}

\author{Gulnisa Abdukayumova}
\email{gulnisaabdukayumova@gmail.com}
\affiliation{National Research University TIIAME, Kori Niyoziy 39, Tashkent 100000, Uzbekistan}

\author{Farruh~Atamurotov}
\email{atamurotov@yahoo.com}
\affiliation{Inha University in Tashkent, Ziyolilar 9, Tashkent 100170, Uzbekistan}
\affiliation{Research Center of Astrophysics and Cosmology, Khazar University, 41 Mehseti Street, Baku AZ1096, Azerbaijan}
\affiliation{University of Tashkent for Applied Sciences, Str. Gavhar 1, Tashkent 100149, Uzbekistan}

\author{Ahmadjon Abdujabbarov}
\email{ahmadjon@astrin.uz}
\affiliation{New Uzbekistan University, Movarounnahr street 1, Tashkent 100000, Uzbekistan}
\affiliation{School of Physics, Harbin Institute of Technology, Harbin 150001, People’s Republic of China}

\author{X. Gao}
\email{gaoxl@zjnu.edu.cn}
\affiliation{Department of Physics, Zhejiang Normal University, Jinhua 321004, People's Republic of China}

\author{G. Mustafa}
\email{gmustafa3828@gmail.com}
\affiliation{ Department of Physics, Zhejiang Normal University, Jinhua 321004, People's Republic of China}

\begin{abstract}
We investigate the dynamics of charged and neutral particles in the vicinity of a Schwarzschild-like black hole immersed in an external magnetic field. We find that the innermost stable circular orbits (ISCOs) for charged particles are systematically smaller than those of neutral particles, demonstrating a fundamental distinction in their orbital dynamics. In the presence of a strong magnetic field, charged particle ISCOs can approach arbitrarily close to the event horizon. We show that collisions between charged particles in ISCOs and neutral particles falling from infinity can produce unbounded center-of-mass energies in the strong-field regime, suggesting the black hole magnetosphere as a natural particle accelerator. Additionally, we apply the relativistic precession model to study quasi-periodic oscillations around the Schwarzschild-like black hole, treating orbital perturbations as coupled harmonic oscillators. Our results provide new insights into high-energy astrophysical processes near magnetized black holes and offer observational signatures through quasi-periodic oscillating frequencies that could be detected by current and future X-ray missions.

\end{abstract}

\maketitle

\section{Introduction}

According to the no-hair theorem, black holes may not have their own magnetic field ~\cite{Misner73}. Astrophysical black holes may have only mass, spin, and electric charge. However, the electromagnetic field plays an important role in astrophysical processes that involve compact objects. In particular, the jet formation from the compact object is followed by the magnetic field due to the accretion disc~\cite{1977MNRAS.179..433B}. A black hole can be considered as embedded in the external electromagnetic field. Wald~\cite{Wald1974PRD} had considered the rotating Kerr black hole immersed in an external asymptotically uniform magnetic field and obtained the solution of the Maxwell equation. It has also been shown that the space-time curvature will change the original structure of the uniform magnetic field. The solution of the Maxwell equation in the case of the current loop surrounding the black hole has been obtained by Petterson~\cite{1974PhRvD..10.3166P}.  The detailed study of the specifications of the change in magnetic field structure and the motion of charged test particles around a compact object in the presence of a magnetic field in Refs.~ \cite{1975PhRvD..12.2218P,2004CQGra..21.3433P,2008GReGr..40.2515A,2013PhRvD..87l5003T,2019PhRvD..99j4009N,2023EPJC...83..854V,Jawad16,Hussain15,Jamil15,Hussain17,DeLaurentis2018PhRvD,Narzilloev20a}.
The magnetic field structure surrounding compact objects may be affected by the effects of the parameters of alternative or modified theories of gravity~\cite{Kolos17,Kovar14,Aliev89,Aliev02,Shaymatov14,Abdujabbarov10,Abdujabbarov11a,Stuchlik16,Turimov18b,Shaymatov20egb,2021PDU....3400891S}.
The study of dynamics of particles with non-zero magnetic dipole momentum around black hole in the presence of an external magnetic field can be found in Refs.~\cite{deFelice,defelice2004,Toshmatov15d,Rahimov11a,Davlataliev:2024wdd,Narzilloev20c,Rayimbaev20d,Vrba20,Abdujabbarov20,Davlataliev:2024smq}.

There is a belief that a black hole is the source of the most energetic objects in the universe, such as active galactic nuclei, ultra-luminous X-ray binaries, gamma-ray bursts, etc. The leading dynamo of the energetics of these objects is the gravity surrounding the black holes. Some of the energy extraction mechanisms from black holes are based on the matter accretion onto the central object. Another way is to extract the thermodynamical energy of the black hole (which is also defined through the gravitational energy of the compact object). In particular, the thermodynamics of the spacetime surrounding the compact objects in different gravity models have been explored in Refs.~\cite{rakhimova2023thermodynamical,mustafa2024probing,ladghami2024barrow,wu2024thermodynamical,ditta2024thermal}.    

The rotational energy of the Kerr black hole may be extracted through the so-called Penrose process~\cite{penrose1969gravitational}. According to this mechanism, particle fission that takes place in the ergosphere may cause particle creation with negative energy at infinity, and escaping the latter corresponds to energy extraction. Physically, it may be interpreted as an extraction of the kinetic energy of a rotating black hole. The magnetic field surrounding the rotating black hole may also affect the Penrose process. The Penrose process for charged particles in the presence of an external magnetic field is referred to as the Magnetic Penrose Process (MPP) and has been extensively studied in Refs.~\cite{PRP4,PRP5, wagh1989energetics,PRP2}. An alternative scenario where the dipole magnetic field generated by the accretion disc around the black hole plays a role in the source of relativistic jet formation is proposed by Blandford and Znajeck.  

Black holes may also play the role of a particle accelerator. Banados, Silk, and West have studied the center of mass energy of the particles falling into rotating black holes and observed that the latter may reach ultra-high values~\cite{Banados09}. Moreover, it has been shown that for the fine-tuned values of the angular momentum of the particles, the center of mass energy of colliding particles at the horizon of an extremely rotating black hole may diverge~\cite{Banados09}. The center of mass energy and energetic processes of the particles around black holes in different gravity models have been extensively studied in Refs.~\cite{2020PhLB..81035850H,2019PhRvD.100b4050Z,2017PhRvD..96j4050S,2023EPJP..138.1022T,2022PhRvD.105b4014H,2021JCAP...08..045A,2021PhRvD.103h4057B,2024NuPhB100516583N,Davlataliev:2024ekv,2023EPJP..138..846N,2015EPJC...75..399A}. The magnetic field may also affect the acceleration process of charged particles near black holes~\cite{Frolov10,Frolov11,Frolov12,Frolov12b}. The magnetic field may increase the center of mass energy of the particles near the static black holes and mimic the effect of rotation~\cite{Tursunov13,Tursunov13a}.

Special attention is given to specific cases of particle dynamics, including circular orbits, as they help us to understand the various astrophysical phenomena. These include accretion disks and the energy of X-ray sources surrounding black holes and compact objects. Understanding the phenomena known as quasi-periodic oscillations (QPOs) may also result from it. By examining the power spectra of the flux from X-ray binary pulsars, it was discovered for the first time. The recent observational data on QPOs can also be considered to constrain the parameters in the framework of gravity models, as explored in Refs.~\cite{2023EPJC...83..572R,2024PhRvD.109j4074A,2024EPJC...84..420A,2024PDU....4601569D,2024JHEAp..43...51D}.

Regular black holes with asymptotically Minkowski cores refer to the theoretical black hole constructs that avoid singularities by incorporating a core where spacetime transitions seamlessly to a Minkowski-like or flat geometry \cite{Simpson:2019mud}. In contrast to conventional black holes with singular centers, these models are predicated on modified gravity or nonlinear electrodynamics, guaranteeing finite curvature and physical quantities. Such solutions aspire to harmonize general relativity with quantum gravity principles, providing a more comprehensive depiction of black hole interiors while maintaining classical horizons. The aim of this research is to investigate the dynamics of neutral and charged particles with dipole moments in the vicinity of Schwarzschild-like black holes~\cite{Simpson:2019mud,Simpson:2021dyo}. 
In addition, we will examine the interactions between these particles and the surrounding field and the collisions that occur. 
{A novel solution for the metric tensor that describes a regular black hole of Bardeen type has been proposed in~\cite{2020Univ....7....2B}. This solution has an asymptotically Minkowski core, and the energy density and pressures vanish at infinity.}
The outline of the paper is as follows:
\begin{itemize}
    \item Section \ref{sec2}: We explore geodesic motion and analyze the circular motion exhibited by neutral massive particles. 
\end{itemize}

\begin{itemize}
    \item Section \ref{sec3}: Here we provide the expression for the magnetic field around a black hole and derive the equations of motion governing the trajectory of a charged particle.
\end{itemize}

\begin{itemize}
    \item Section \ref{sec4}: Here we present the dynamics of a particle in a dimensionless form to study the characteristic features of its motion and to discuss the collisions of particles in the vicinity of a black hole.
\end{itemize}

\begin{itemize}
\item Section \ref{qpo}: The fundamental frequencies related to the circular orbits of the particles were examined. The application of particle motion and fundamental frequencies to QPO analysis was discussed.
\end{itemize}

\begin{itemize}
    \item Section \ref{Sec:conclusions}: Finally, we summarize the main conclusions and results obtained during the research.
\end{itemize}

Throughout this paper, we will use the metric signature of $(-, +, +, +)$ for the spacetime metric and a system of units in which the gravitational constant $(G)$ and the speed of light $(c)$ are set to $1$. Greek indices will range from $0$ to $3$, while the Latin indices will range from $1$ to $3$.

\section{Geodesic Motion}
\label{sec2}
We consider electromagnetic ﬁelds of compact astrophysical objects in Schwarzschild-like spacetime, which in a spherical symmetric Boyer-Linquist coordinate system ($t$, $r$, $\theta$, $\phi$) is described by the metric \cite{Simpson:2019mud,2022PhRvD.105f4065S,Culetu:2014lca, Culetu:2015cna}:
\begin{equation}
ds^2=-f(r)dt^2+\frac{1}{f(r)}dr^2+r^2d\theta^2+r^2\sin^2{\theta}d\phi^2,
\label{eq: 1}
\end{equation}
where
\begin{equation}
f(r)=1-\frac{2Me^{-a/r}}{r}.
\end{equation}

Also, when $a = 0$, the metric (\ref{eq: 1}) expresses the pure Schwarzschild spacetime.

\begin{figure}[H]
\centering
\includegraphics[width=1\linewidth]{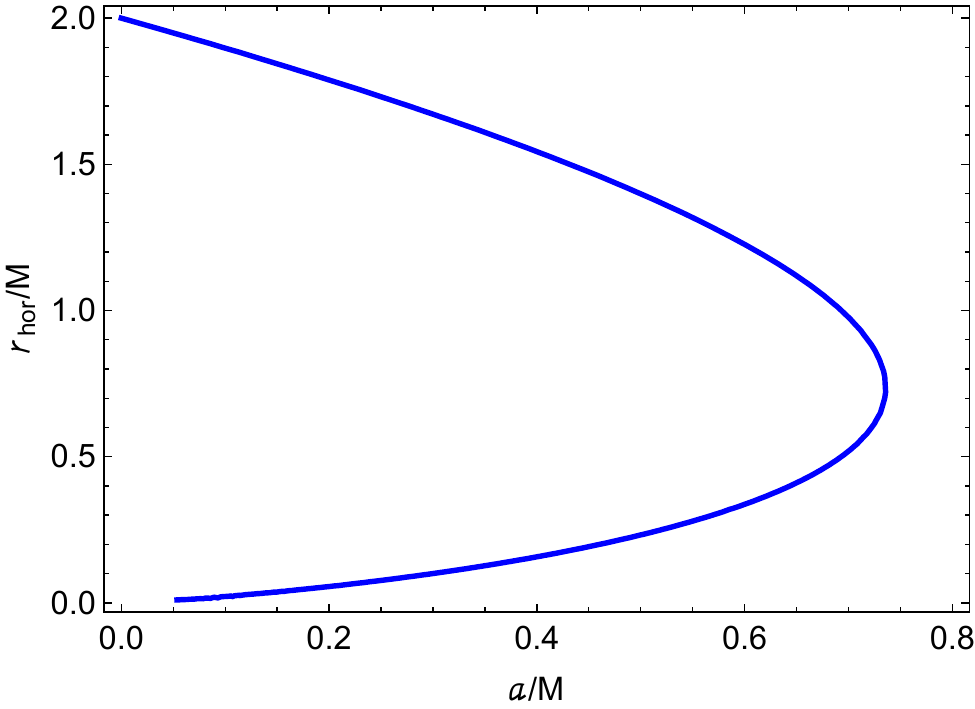}
\caption{Dependence of horizon radius from matric parameter $a$ in Schwarzschild-like spacetime.}
\label{fig:enter-label}
\label{1}
\end{figure}

The event horizon of the black hole is positioned at the real root where $g^{rr} = 0$, and its variation with the metric parameters is illustrated in Fig.~\ref{1}. 
\begin{equation}
    r_{h}=-\frac{a}{Productlog(-a/2M)}.
    \label{horizon}
\end{equation}
Where $Productlog$ gives the principal solution for $w$ in $z=we^w$, that is, the Lambert function, the inverse function of the exponent~\cite{Corless:1996zz}.
In Fig.~\ref{1} shows the dependence of the event horizon. As we can see, our Schwarzschild-like metric also turns out to have two event horizons. But the parameter $a$ in our metric does not reach 1, so in the further study we use $a/M<1$.

Now, we consider the dynamics of a neutral particle in a Schwarzschild-like spacetime expressed by Eq.~(\ref{eq: 1}). There are three constants of motion corresponding Eq.~(\ref{eq: 1}) in which two of them originate as a result of two Killing vectors: 
\begin{equation}
\label{eq:5} 
\xi_{(t)} = \xi^{\mu}_{(t)} = \partial_t, \ \
\xi_{(\phi)} = \xi^{\mu}_{(\phi)} = \partial_{\phi},
\end{equation}
where $\xi^\mu_{(t)}=(1,0,0,0)$ and $\xi^\mu_{(\phi)}=(0,0,0,1)$. Eq.~(\ref{eq:5}) indicates that the metric describing the black hole remains unchanged when subjected to shifts in time and rotations about the axis of symmetry, respectfully. 

The equation of equatorial motion for a neutral particle with specific energy \(\mathcal{E} = {E}/{m}\) and specific angular momentum \(\mathcal{L} = {L}/{m}\) can be represented as
\begin{align}
    &\dot{t}=\frac{\mathcal{E}}{f}, \quad \dot{\phi}=\frac{\mathcal{L}}{r^2} , \\
    &\dot{r}^2=\mathcal{E}^2-f\left(\epsilon+\frac{\mathcal{L}^2}{r^2}\right), 
\end{align}
where the dot signifies the differentiation operation with respect to an affine parameter along the particle’s trajectory, while the parameter $\epsilon$ determines whether the geodesic is null ($\epsilon=0$) or timelike ($\epsilon=1$).

\subsection{The circular orbits}
\begin{figure}[H]
     \centering
     \includegraphics[width=1\linewidth]{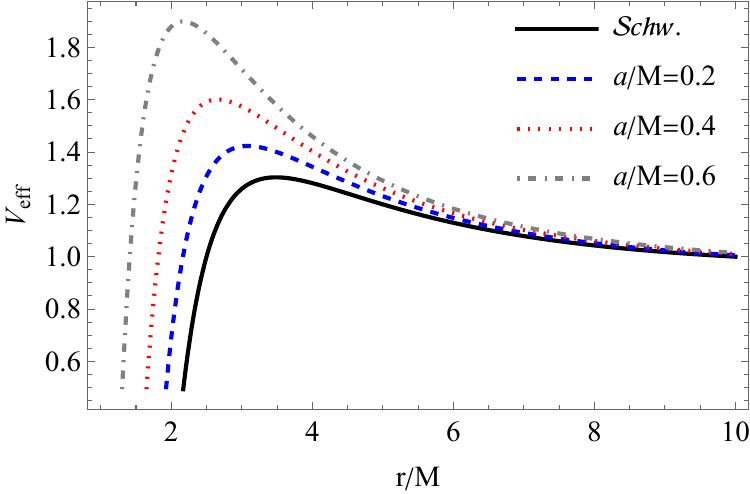}
    \caption{The radial dependence of effective potential of neutral particle as a function of $r/M$ for different values of parameter $a$ ($\mathcal{L}/M =5$).} 
    \label{fig2}
 \end{figure}

To investigate a specific class of orbits for neutral particles, such as innermost stable circular orbits (ISCOs) and marginally bound orbits (MBO), we consider the motion of particles with \(\epsilon = 1\). In this case, the radial equation for a massive particle can be written as follows.

\begin{align}
&\dot{r}^2=\mathcal{E}^2-V_{eff}, \nonumber\\
&V_{eff}=\left(1-\frac{2Me^{-a/r}}{r}\right)\left(1+\frac{\mathcal{L}^2}{r^2}\right). 
\end{align}
Figure~\ref{fig2} shows the radial dependence of the effective potential for the circular motion of neutral particles around Schwarzschild-like black holes for various values of the parameters $a$ and \(\mathcal{L}\). In particular, the maximum can increase or decrease depending on the value of \(a\). In addition, an increase in the value of the parameter \(\mathcal{L}\) leads to a significant decrease in the minimum of the effective potential.

When considering conditions where $\dot{r}=\ddot{r}=0$, indicating the absence of radial motion and radial acceleration of the particle around the object, one can derive the critical specific energy and specific angular momentum of the particle.
This revised version clarifies that under these conditions, the specific energy and angular momentum can be determined as:

\begin{eqnarray}
    \mathcal{E}^2 &=& \frac{f^2 r^2 }{f r^2 - e^{-a/r} M (r -a)}, 
\\
\mathcal{L}^2 &=& \frac{e^{- \frac{a}{r}} M (r-a)}{f r^2 - e^{-a/r} M (r -a)}. 
\end{eqnarray}

\begin{figure}[H]
    \centering
    \includegraphics[width=1\linewidth]{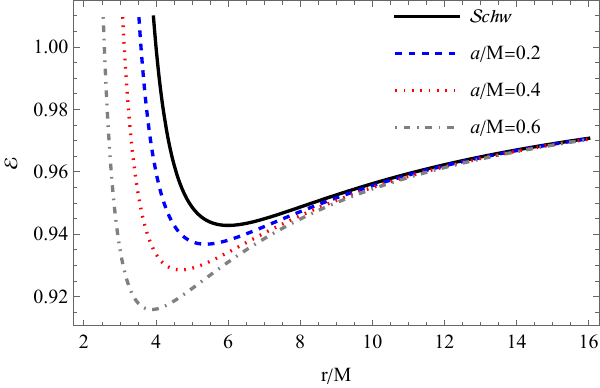}
    \caption{The Radial dependence of energy per mass $\mathcal{E} $ of neutral particle
orbiting around the Schwarzschild-like black hole for the different values of  parameter $a$.}
   \label{Fig3}
\end{figure}

\begin{figure}[H]
    \centering
    \includegraphics[width=0.9\linewidth]{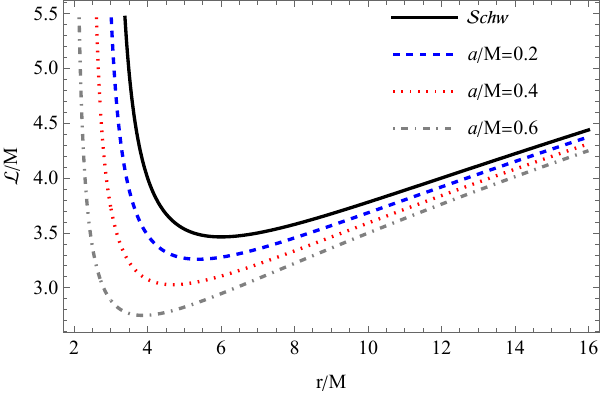}
    \caption{The Radial dependence of the angular momentum per mass $\mathcal{L}/M$ of a neutral particle orbiting around the Schwarzschild-like black hole for the different values of  parameter $a$.}
    \label{Fig4}
\end{figure}

The radial dependence of these expressions is shown in Figs.~\ref{Fig3} and~\ref{Fig4}.
These figures also show that an increase in parameter $a$ leads to a weaker interaction between the particle and the black hole.


\subsection{Orbital velocities}

\begin{figure}[H]
    \centering
   \includegraphics[width=0.89\linewidth]{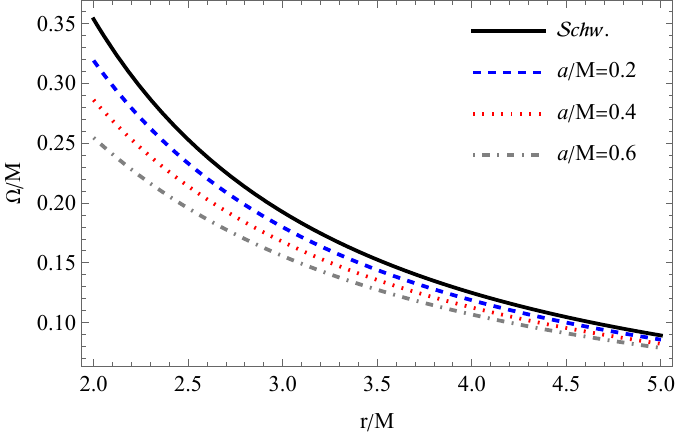}
   \caption{The  Radial dependence of Keplerian angular frequency of neutral particle orbiting around the Schwarzschild-like black hole for the different values of  parameter $a$.}
    \label{fig3}
\end{figure}

\begin{figure}[H]
    \centering
   \includegraphics[width=0.89\linewidth]{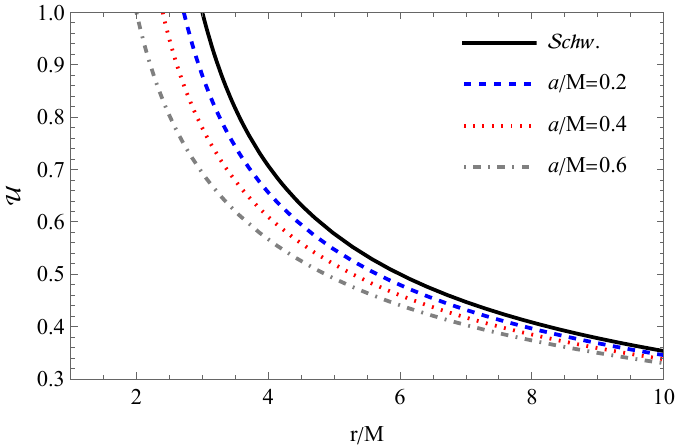}
 \caption{The  Radial dependence of orbital velocity of neutral
particle orbiting around the Schwarzschild-like black hole for the different values of parameter $a$.}
    \label{fig4}
\end{figure}

Now, we can study the angular and linear velocities of a neutral particle orbiting a black hole. The angular velocity, which describes how fast an object rotates around a central point, can be represented as~\cite{Boboqambarova:2021cbf}:
\begin{equation}
    \Omega = \sqrt{- \frac{\partial_{r} g_{tt}}{\partial_{r} g_{\phi \phi}}} = \frac{\sqrt{e^{- \frac{a}{r}} M (r-a)}}{r^2}. 
\end{equation}
The dependence of angular velocity on radius is shown in Fig.~\ref{fig3}. Here, we can also check its value for the Schwarzschild metric, which is appropriate for $a = 0$; the angular velocity will be equal to $\Omega=\sqrt{M/r^3}$.
The required orbital velocity, which is the velocity that balances the inward gravitational force with the outward centrifugal force, allowing an object to follow a circular or elliptical path around the central body, measured by a local observer depends, which is illustrated in Fig.~\ref{fig4}, on the distance from the black hole and its mass in our metric as:
\begin{equation}
\mathcal{U} = \sqrt{\frac{g_{\phi \phi} \, \partial_{r} g_{tt}}{g_{tt} \, \partial_{r} g_{\phi \phi}}} = \sqrt{ \frac{M (1 - a/r)}{r (e^{-a/r} - 2M / r) }}.
\end{equation}
The metric parameter $a$ decreases the angular frequency and the orbital velocity, as illustrated in Figs.~\ref{fig3} and \ref{fig4}.

\section{Charged particle dynamics}
\label{sec3}
\subsection{Magnetic field configuration}
In realistic astrophysical scenarios, the magnetic field structures near black holes are highly intricate because of the behavior of ionized matter in the accretion disk. A straightforward method to derive an analytical expression for the non-zero components of an external magnetic field around a Schwarzschild black hole is the approach \cite{Wald1974PRD}. This provides the following precise analytical expression for the electromagnetic vector potential 
\begin{equation}
A_\mu = \frac{B}{2} \xi_{\mu(\phi)} .
\end{equation}
Using Eq.~(\ref{eq:5}), we can find nonzero components of 4-potential:
\begin{equation}
A_{\phi (Schw.)}=\frac{B}{2}r^2\sin^2{\theta}. 
\label{eq: 3}  \end{equation}
In curved spacetime, Maxwell's equations take
the form:
\begin{equation}
   \frac{1}{\sqrt{-g}} \, \frac{\partial }{\partial x^\nu} \left(\sqrt{-g} \, F^{\mu \nu}  \right) = 0, \quad F_{\mu \nu} = A_{\nu , \mu} - A_{\mu , \nu} .
   \label{eq: 2}
\end{equation}

Now Eq.~(\ref{eq: 2}) becomes:
\begin{equation}\label{eqmax}
    \frac{1}{\sin\theta} \, \frac{\partial}{\partial r} (f A_{\phi , r}) + \frac{1}{r^2} \frac{\partial}{\partial \theta} \left( \frac{1}{\sin\theta} A_{\phi , \theta}\right) = 0 .
\end{equation}

In a similar manner, within the context of the spacetime metric given by Eq.~(\ref{eq: 1}), the expression for the vector potential can be divided as follows:

\begin{equation}
    A_{\phi} = \frac{1}{2} B \, \psi(r) \sin^2\theta, 
    \label{2}
\end{equation}
where $\psi(r)$ is radial function found as a solution of Maxwell Eq.~(\ref{eq: 2}).

Substituting Eq.~(\ref{2}) into Eq.~(\ref{eqmax}), one can obtain:

\begin{equation}
    r^2 (f \psi')' - 2 \psi = 0, 
    \label{12}
\end{equation}
where a prime indicates the derivative with respect to the radial coordinate; finally, the precise analytical solution of Maxwell's equation for the vector potential near a Schwarzschild-like black hole can be determined by solving the differential Eq.~(\ref{12}). Furthermore, employing the Taylor series expansion, we obtained the component of the 4-potential: 

\begin{equation}
    A_{\phi} = \frac{1}{2} \, B \, (r^2 - 2 M a) \, \sin^2\theta. 
\end{equation}
Now, considering the nonzero components of the 4-potential, we determine the components of the electromagnetic field tensor as follows:

\begin{eqnarray}
    F^{r \phi} = -F^{\phi r} &=& g^{r r} g^{\phi \phi} A_{\phi , r} = f \frac{B\sin{\theta}}{r} ,
\\
    F^{\theta \phi} = -F^{\phi \theta} &=& g^{\theta \theta} g^{\phi \phi} A_{\phi , \theta} \nonumber \\ &=& \frac{B \cos{\theta}}{r^2 \sin{\theta}} - \frac{2 a B M \cos\theta}{r^4 \sin\theta} , 
\end{eqnarray}
where determinant of the metric tensor $\Delta\, g = r^2 \sin\theta$. 
The components of the magnetic field in an orthonormal coordinate system can be expressed using the electromagnetic field tensor as follows:
\begin{eqnarray}
    B^{\hat{r}} &=& B (1 + \frac{2 M a}{r^2}) \cos\theta, 
\\
    B^{\hat{\theta}} &=& B \sqrt{f} \sin\theta. 
\end{eqnarray}

\begin{figure*}[ht]
    \centering
    \subfloat[$a/M = 0$]{\includegraphics[width=0.42\textwidth]{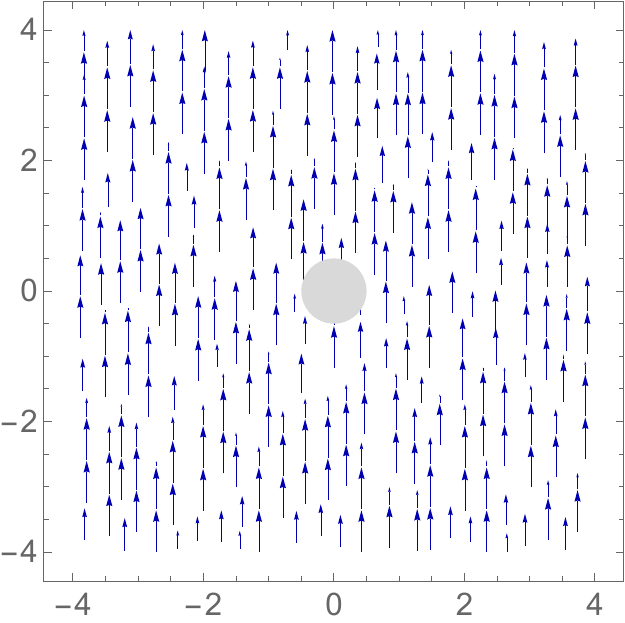}}
    \hfill
    \subfloat[$a/M = 0.2$]{\includegraphics[width=0.42\textwidth]{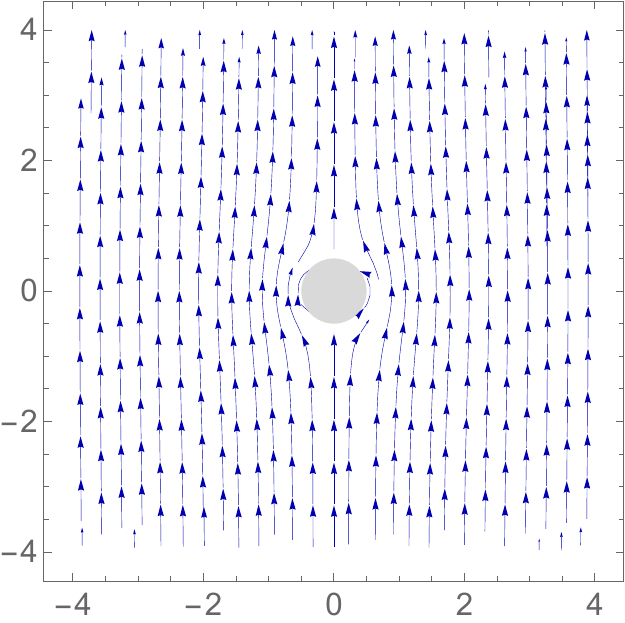}}
    \caption{The magnetic field lines near a Schwarzschild-like black hole for various $a$ parameter values.}
    \label{fig:multiple_graphs}
\end{figure*}

To investigate the magnetic field configuration near the black hole, we can examine the magnetic field lines characterized by the equation \( A_{\phi} = \text{const}\). Fig.~\ref{fig:multiple_graphs} illustrates the magnetic field lines around the Schwarzschild-like black hole for different values of the \(a\) parameter. 

\subsection{Charged particle motion}

Now, one can analyze the motion of a test charged particle around a Schwarzschild-like black hole immersed in an external magnetic field. The Lagrangian for a particle with mass $m$ and charge $q$ can be written as:
\begin{equation}
\mathscr{L}=(x^\mu,\dot{x}^\mu)=\frac{1}{2}g_{\mu\nu}\dot{x}^{\mu}\dot{x}^{\nu}+\frac{q}{m}A_\mu\dot{x}^\mu, 
\label{eqmot}
\end{equation}
where $\dot{x}^\mu=dx^\mu/dS$.
Since we focus primarily on the equatorial motion of a charged particle, it is clear that the given gauge field vanishes in that plane (\(\theta = \pi/2\)). Using Eq.~(\ref{eqmot}), we obtain :
\begin{equation}
    g^{\mu\nu}\left(\frac{\partial S}{\partial x^\mu}-qA_\mu\right)\left(\frac{\partial S}{\partial x^\nu}-qA_\nu\right)=2\frac{\partial S}{\partial \tau}. 
\end{equation}
Consequently, the motion of charged particles is governed by the following action:
\begin{equation}
    S=-\frac{1}{2}m^2\tau-\mathcal{E}t+\mathcal{L}\phi+S_{r,\theta}, 
\end{equation}
which arises from the symmetry of the system. The conserved quantities are the specific energy $\mathcal{E} = E/m$ of a moving particle and its angular momentum $\mathcal{L} = L/m$, respectively, and the corresponding function \( S_{r,\theta} \) are dependent on \( r \) and \( \theta \). Now, it is simple to distinguish between the variables in the equatorial plane, allowing for the derivation of the equation for radial motion:
\begin{align}
\dot{r}^2=\mathcal{E}^2-f-\frac{f}{r^2}\left(\mathcal{L}-\frac{q A_{\phi}}{m}\right)^2 = \, \mathcal{E}^2-V_{eff}(r), 
\end{align}
the effective potential $V_{eff}(r)$ represents the circular motion of test particles and is given by the following expression:
\begin{equation}\label{eqvef}
V_{eff}=f\left[1+\frac{\left(\mathcal{L}-\frac{qB}{2m}(r^2-2Ma)\right)^2}{r^2}\right]. 
\end{equation}

One can introduce a  magnetic parameter, describing the interaction between charge and the external magnetic field, in the following form:
\begin{equation}
    b=\frac{qB}{2m}. 
    \label{eq: 69}
\end{equation}
Now, it is possible to determine how magnetic and metric parameters influence the motion of charged particles. The magnetic parameter modifies the shape of the effective potential, shifting it toward the central object. Consequently, the minimum possible distance between charged particles and the black hole decreases.

\begin{figure*}[ht]
    \centering
   {\includegraphics[width=0.45\textwidth]{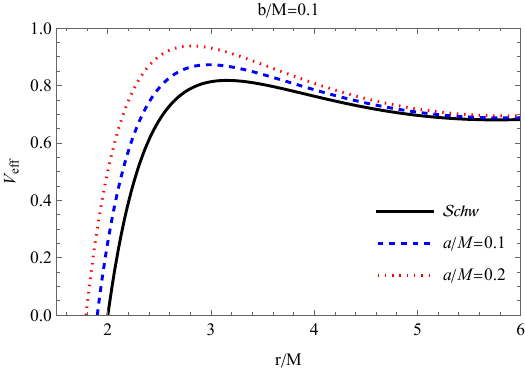}}\hfill
 {\includegraphics[width=0.45\textwidth]{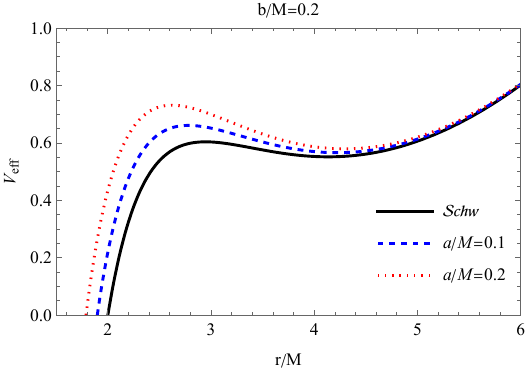}}\hfill
    \caption{\justifying The radial dependence of the effective potential $V_{eff}$ of a charged particle for different values of the metric parameter $a$ for a fixed magnetic parameter $b$. }
    \label{fig:subfig1}
\end{figure*}

As shown in Fig.~\ref{fig:subfig1}, the parameter $a$ affects the structure of the effective potential in the vicinity of the central object. An increase in $a$ causes the effective potential to shift inward, leading to a reduction in the radius of circular orbits for test particles.

Our analysis indicates that in the presence of an external magnetic field, the effective potential diverges at large distances:

\begin{equation}
 \lim_{r \to \infty} V_{eff} = \infty. 
\end{equation}
The previous section noted that the magnetic field solution was derived near the black hole, enabling us to reliably use the effective potential in Eq.~(\ref{eq: 69}) for the orbit of a charged particle close to the black hole.

\subsection{Stable Circular Orbits}

To describe the stable circular orbits of  particles around the central black hole, we employ the following standard conditions \cite{Narzilloev:2020qtd}:
\begin{equation}
    V_{eff} = \mathcal{E}^2, \quad V_{eff}' = 0, \quad V_{eff}'' \geq 0. 
    \label{eq: 7}
\end{equation}

From these expressions, we can find the radial dependence of the energy and the angular momentum. 

\begin{figure*}[ht]
    \centering
    \subfloat{\includegraphics[width=0.45\textwidth]{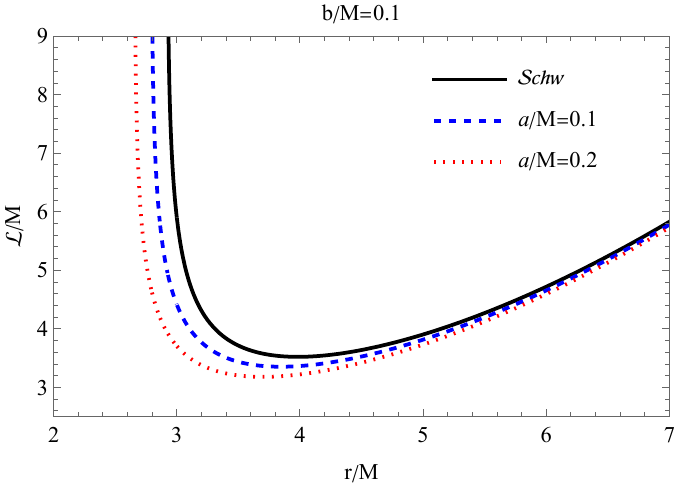}}\hfill
     \subfloat{\includegraphics[width=0.45\textwidth]{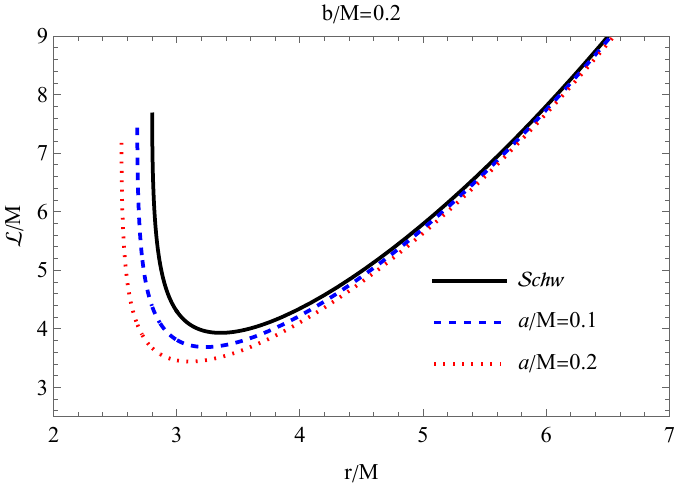}}\hfill
\caption{\justifying Radial dependence of the angular momentum per mass $\mathcal{L}/M$ of radial motion of charged particle for different values of magnetic parameter $b$. }
\label{Fig.L(b)}
\end{figure*}
\begin{figure*}[ht]
    \centering
    \subfloat{\includegraphics[width=0.45\textwidth]{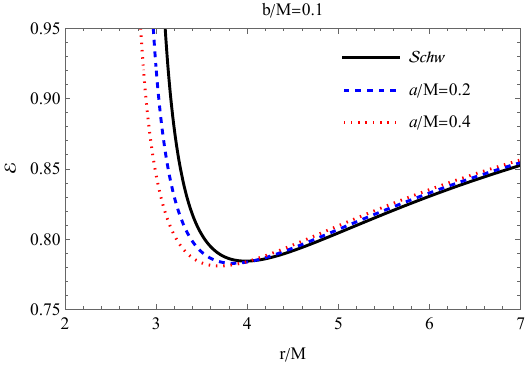}}\hfill
    \subfloat{\includegraphics[width=0.45\textwidth]{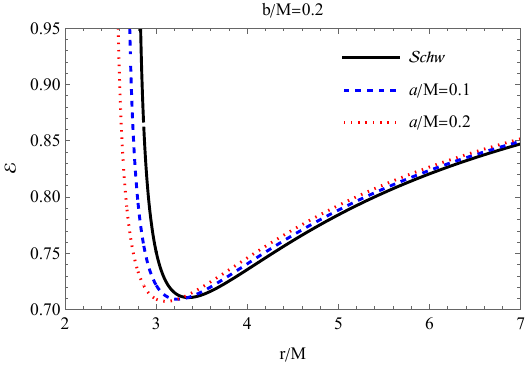}}\hfill
    \caption{\justifying Radial dependence of the energy $\mathcal{E}$ of radial
motion of charged particle for different values of magnetic parameter $b$.  }
\label{Fig.E(b)}
\end{figure*}
\begin{figure}[t]
    \centering
    {\includegraphics[width=0.46\textwidth]{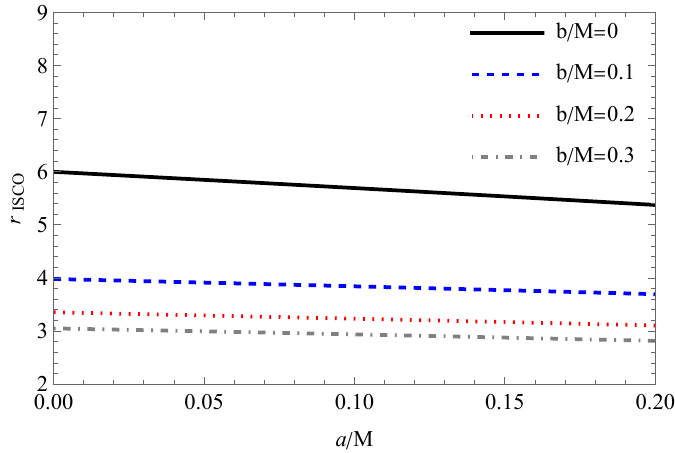}}
    \caption{\justifying Dependence of the ISCOs position for a charged particle orbiting around the Schwarzschild-like black hole from the  parameter $a$ for the fixed $b$ values.  }
\label{mob}
\end{figure}

Figures~\ref{Fig.L(b)} and ~\ref{Fig.E(b)} demonstrate the radial dependencies of the angular momentum and energy of a particle for fixed values of the black hole parameter $a$ and the magnetic parameter $b$, respectfully. An increase in the parameter $b$ leads to a decrease in the energy value and an increase in the angular momentum. Also, to study the effect of the magnetic parameter on the motion of the particle, we have given a graphical dependence of the radius of the ISCOs in Fig.~\ref{mob}. As can be seen in the graphs, the ISCOs decrease as the parameter $b$ increases. This suggests that the magnetic field assists in maintaining the particle in a stable orbit around the black hole at lower energies and distances.


\section{Particle collision}
\label{sec4}
\subsection{Dimensionless form of dynamical equations}\label{sec6}
In this chapter we discuss different cases of collision of charged particles near the black hole. We are particularly interested in the possibility that particles are accelerated by a black hole. 
For this, first we will transform and derive the equations of motion obtained in sections \ref{sec2} and \ref{sec3} in dimensionless form. For the purpose of further research and to simplify the expression, let us assume the value of $M = 1$.
Since our Schwarzschild-like metric is given in the form Eq.~(\ref{eq: 1}) and because of the presence of two Killing vectors in Eq.~(\ref{eq:5}), we can write down the constants of motion using the 4-momentum in the following form:
\begin{align}
   & E=-\xi^{\mu}_{(t)} p_\mu=m \frac{d t}{d \tau}f, \label{eq:38}\\
    &L=\xi^{\mu}_{\phi}p_{\mu}=\left[m \frac{d \phi}{d \tau}r^2+\frac{1}{2}q B (r^2-2 a) \right]\sin^2{\theta}, 
\end{align}
where $p_\mu$ is the generalization of 4-momentum.
For more convenient use, we use the following dimensionless versions of $r,\mathcal{L}, \tau, b $:

\begin{align}
    \rho=\frac{r}{r_h e^a},\quad \sigma=\frac{\tau}{r_h e^a},\quad l=\frac{L}{m r_h e^a},\nonumber \\
   \mathcal{E}=\frac{E}{m}, \quad b=\frac{q B r_h e^a}{2 m}, \quad f=1-\frac{e^{-a}}{\rho}, \nonumber 
\end{align}
Where $r_h$ is the radius of the horizon of our metric from Eq.~(\ref{horizon}) . We use the expression for 4-momentum of the particle $u^{\mu} u_{\mu}=-1$, and obtain the resulting dimensionless equations:
\begin{eqnarray}
 &&   \left(\frac{d \rho}{d \sigma}\right)^2= \mathcal{E}^2 - U, \\ && \frac{d t}{d \sigma}=\frac{\mathcal{E} \rho}{\rho-e^a},\\ && \rho\frac{d \phi}{d \sigma}=\beta,\quad  \beta=\frac{1}{\rho}\left(l+\frac{2ab}{(r_h e^a)^2}\right)-b \rho, 
\end{eqnarray}
where $U$  is the effective potential given by the equation:
\begin{equation}
    U=\left( 1-\frac{e^{-a}}{\rho}\right)(1+\beta^2). 
\end{equation}

The equations $U_{,\rho} = 0$ and $U_{,\rho \rho} = 0$ define the ISCOs. By solving this system of equations, we obtain an explicit expression for $l$ and $b$:
\begin{equation}
b^2_{\pm} = \frac{9 \mp N -e^a \rho (4 e^{2a}\rho^2-21e^a\rho+30)}{8 \, \rho^2 \, (e^{a} \, \rho - 1)^2 \, (4 \, e^{2a} \, \rho^2 - 10 \, e^{a} \, \rho + 3)}
\label{eq:49}
\end{equation}
\begin{align}
&l_{\pm} = b_{\pm} \, e^{-2a}\,\frac{e^{2a} \, \rho^2 \,r_h^2\, N\pm 2a \, (e^a\rho-3)^2 }{(e^a \, \rho-3)^2 \, r_h^2}.
\label{eq:50}
\end{align}
where 
\begin{equation}
    N = \sqrt{\left(1-3 e^a \rho \right) \left(e^a \rho -3\right)^3}
\end{equation}

In the case of Schwarzschild spacetime, the equation exists within the interval $\rho \in (1/3, \, 3]$.

\subsection{Two particles at ISCOs}
We study the energy of particles that collide around a black hole in the Schwarzschild-like metric using Frolov's method \cite{Frolov2012}.

It is well-known that particles moving along ISCOs have a minimum magnitude of energy and angular momentum, and their velocities become close to the light speed. Because of this, we get the nonzero components of the momentum of the particles found in the form:
\begin{eqnarray}
\label{29}
p^t&=&m\gamma e^t_{(t)}, 
    \\
    \label{30}
         p^\phi&=&m \gamma v e^\phi_{(\phi)}, 
   \\
    e^t_{(t)}&=&\sqrt{\eta_{tt} g^{tt}}=\frac{1}{\sqrt{f}} , 
    \label{eq: 1224}
    \\
     e^\phi_{(\phi)}&=&\sqrt{\eta_{\phi \phi} g^{\phi \phi}}=\frac{1}{r}, 
    \label{eq: 1223}
\end{eqnarray}
where Eq.~(\ref{eq: 1224}) and Eq.~(\ref{eq: 1223}) are orthonormal tetrad components in equatorial motion around fixed orbit. Here $v$ (which can be both positive and negative) is a
velocity of the particle with respect to a rest frame, and $\gamma$ is the Lorentz gamma factor:
\begin{equation}
    \frac{d \phi}{d \tau}=\frac{p^\phi}{m}=\frac{v \gamma}{r}. 
\end{equation}
From Eq.~(\ref{eq:38}) one may easily get the following 
\begin{equation}
      \frac{d \phi}{d \tau}=\frac{\beta}{r} , \quad         \gamma=\sqrt{1+\beta^2} \, . 
\end{equation}
Here we can use expressions (\ref{eq:49}), (\ref{eq:50}) and rewrite $\gamma$ as:
\begin{equation}
    \gamma_{\pm}=\sqrt{\frac{-9\pm N+e^a\rho(4 \, e^{2a} \, \rho^2 - 21 \, e^{a} \, \rho + 30)}{(e^a\rho-3)(4 \, e^{2a} \, \rho^2 - 10 \, e^{a} \, \rho + 3)}}
\end{equation}

\begin{figure}[H]
    \centering
    \includegraphics[width=0.9\linewidth]{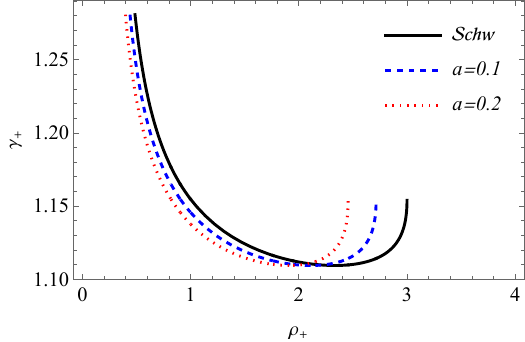}
    \caption{The value of the $\gamma$ factor $\gamma_{+}$ at the position of the ISCOs $\rho_{+}$. }
    \label{gamma+}
\end{figure}

\begin{figure}[H]
    \centering
    \includegraphics[width=0.9\linewidth]{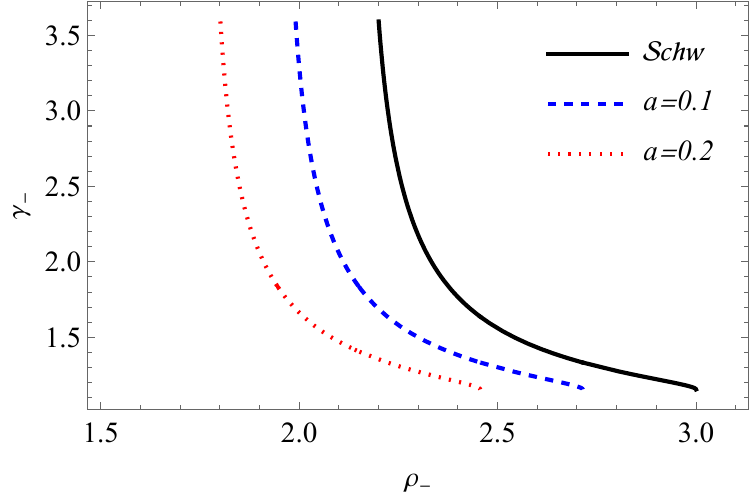}
    \caption{The value of the $\gamma$ factor $\gamma_{-}$ at the position
of the ISCOs $\rho_{-}$. }
    \label{gamma-}
\end{figure}

The dependence of the gamma factor on $\rho$ for fixed values of $a$ is shown in Figs.~\ref{gamma+} and \ref{gamma-}. For the ISCOs radius, in the case of the first solution, $\gamma_+$ corresponds to $\rho_+$, and $\gamma_-$ corresponds to the radius $\rho_-$.

As we can see in Fig.~\ref{gamma+}, the gamma factor accepts values slightly more than 1, but in Fig.~\ref{gamma-} $\gamma_{-}$ it obtains quite large values. For $\gamma_+$, the energy cannot be high, as $\gamma_+$ is not significantly different from 1. However, for $\gamma_-$, the energy increases significantly. It can be seen from Figs.~\ref{gamma-} and \ref{gamma+} that the black hole parameter $a$ helps to reduce the radius, while almost not affecting the values of $\gamma$. For a more complete presentation, in Tab.~\ref{tab} the numerical results for the maxima and minima of the quantities of interest to us, the ISCOs radius and the gamma factor, were shown for various values of the parameter $a$.

  \begin{table}[h!]
    \centering
    \begin{tabular}{|m{2.5cm}|m{2.5cm}|m{2.5cm}|}
        \hline
        \multicolumn{1}{|c|}{$a =0$} & \multicolumn{1}{c|}{$a =0.1$} & \multicolumn{1}{c|}{$a =0.2$} \\
        \hline
       $\rho_{+max} = 3$ & $\rho_{+max} = 2.714$ & $\rho_{+max} = 2.456$ \\
        $\rho_{+min} = 0.333$ & $\rho_{+min} = 0.302$ & $\rho_{+min} = 0.273$ \\
        $\rho_{-max} = 3$ &
        $\rho_{-max} = 2.715$ &
        $\rho_{-max} = 2.456$ \\
        $\rho_{-min} = 2.151$ &
        $\rho_{-min} = 1.947$ &
        $\rho_{-min} = 1.761$ \\
        \hline
         $\gamma_{+max} = 1.959$ & $\gamma_{+max} = 1.951$ & $\gamma_{+max} = 1.963$ \\
        $\gamma_{+min} = 1.1541$ & $\gamma_{+min} = 1.154$ & $\gamma_{+min} = 1.154$ \\
        $\gamma_{-max} = 200$ 
        & $\gamma_{-max} = 586$
        & $\gamma_{-max} = 503$ \\
        $\gamma_{-min} = 1.155$ & $\gamma_{-min} = 1.155$ & $\gamma_{-min} = 1.155$ \\
        \hline
    \end{tabular}
    \caption{\justifying Numerical values of extremes $\rho_{\pm}$ and $\gamma_{\pm}$ at fixed values of $a$.}
    \label{tab}
\end{table}
\renewcommand{\arraystretch}{1.2}

As a first example, consider the collision of two particles with the same mass $m$, with charges $+q$, $-q$, and moving along the same circular orbit in opposite directions. The four-momentum of the system after the collision is:
\begin{equation}
    P^0= 2 m \gamma \frac{1}{\sqrt{f}}. 
\end{equation}
The energy after the collision in the center of the mass reference frame:
\begin{equation}
    \mathcal{M}=2m\gamma. 
\end{equation}
 From the analysis given in Tab.~\ref{tab}, it can be seen that in some cases it is possible to observe high energies in a collision corresponding to large gamma values.
 
\subsection{Collision of a freely falling neutral particle
with a charged particle at ISCOs}
\par
Let us now consider another case, when a neutral particle collides with a charged particle revolving at a circular
orbit near a weakly magnetized black hole. We denote the $p$ four-momentum of a charged particle, $m$, and $ q$ its mass and charge. Also, $k$ the four-momentum of neutral particle, $n$ its mass.
\begin{eqnarray}
    P&=&p+k, 
\\    \mathcal{M}^2&=&m^2+n^2-2(p, k), 
\end{eqnarray}
where $(p,k)=g_{\mu \nu} p^{\mu} k^{\nu}$. First, we find an expression for $ k$ using terms of the
integrals of motion for the neutral particle. We use affine parameter to parametrize geodesic motion:
\begin{eqnarray}
    && k=(\dot{t},\dot{r},\dot{\theta},\dot{\phi}) , 
\\
    && g_{\mu \nu} \dot{x}^{\mu} \dot{x}^{\nu}=-n^2. 
\end{eqnarray}
We use the following integrals of motion: the energy  $\mathfrak{E}$, angular momentum $\mathfrak{L}$ and   azimuthal angular momentum $\mathfrak{L}_z$
\begin{equation}
   \mathfrak{E}=f\dot{t},\quad  \mathfrak{L}_z=r^2\sin^2{\theta} \dot{\phi},\quad \mathfrak{L}^2= r^4\dot{\theta}^2+\frac{\mathfrak{L}_z^2}{\sin^2{\theta}}. 
\end{equation}
Based on this, we have:
\begin{align}
& k=(\frac{\mathfrak{E}}{f},\dot{r},\dot{\theta},\frac{\mathfrak{L}_z}{r^2\sin^2{\theta}}), \\
&\dot{\theta}=\pm \frac{1}{r^2}\sqrt{\mathfrak{\L}^2-\frac{\mathfrak{L}_z}{r^2\sin^2{\theta}}}, \\
&\label{eq47} \dot{r}=\pm \sqrt{\mathfrak{E}^2-\left(n^2+\frac{\mathfrak{L}^2}{r^2}\right)f}\ . 
\end{align}

To find the critical value of the angular momentum $\mathfrak{L}$ that distinguishes between capture by a black hole and escape to infinity for a particle starting from infinity, we use the following relationship, reformulated from Eq.~(\ref{eq47}):
\begin{equation}
    \dot{r}^2=\mathfrak{E}^2 W,\quad W=1-\left(\nu^2+\frac{l^2}{\rho^2}\right)\left(1-\frac{e^{-a}}{\rho}\right). 
\end{equation}
Here $\nu=n/(\mathfrak{E})$,\quad $l=\mathfrak{L}/(\mathfrak{E} r_h e^a)$. Analyzing these expressions, we can conclude that $\nu\leq1$ for motion from infinity, $\nu=0$ for ultra-relativistic particles and light. We can find  the critical angular momentum based on the given conditions:
\begin{equation}
    W=0,\quad d W/ d \rho =0\label{Eq49}. 
\end{equation}
Since the condition $W=0$ determines the radial turning points and from the condition $d W/ d \rho =0$, we find the maximum of $W$. From the solving of Eq.~(\ref{Eq49}), we find values of the critical impact parameter $l_{crit}$.

When a particle has an angular momentum $l <  l_{crit} $, it will be captured by the black hole. If $l >  l_{crit} $, the particle will have a close encounter and then escape to infinity.

\begin{figure}[H]
    \centering
    \includegraphics[width=1\linewidth]{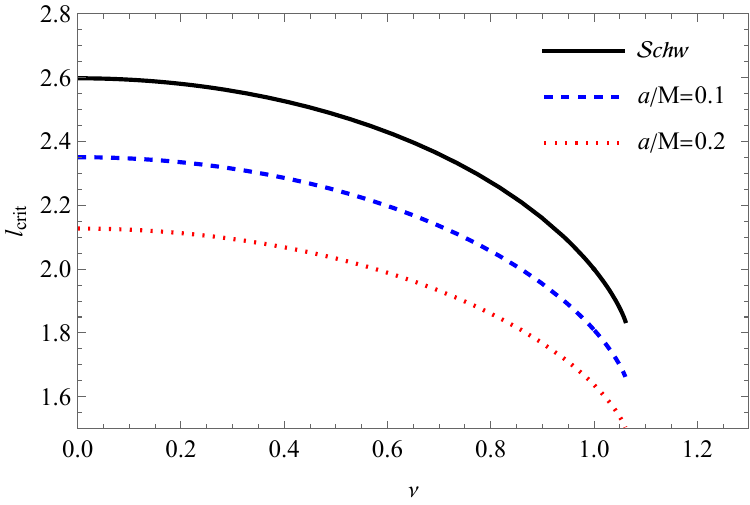}
    \caption{Critical angular momentum as a function of $\nu=n/(\mathfrak{E})$ for fixed values of $a$.}
    \label{fig13}
\end{figure}

This critical value $l_{crit}$ depends on the specifics of the black hole's mass and the initial conditions of the particle. Its dependence on the parameter $\nu$ is shown in Fig.~\ref{fig13}
Here, because of the similar structure, the graph of the dependence of the critical angular momentum in our metric coincides with the graph for the Schwarzschild metric.
With this representation of $k$ and Eq.~\eqref{29}, Eq.~\eqref{30}, we find
\begin{equation}
    (p,k)=-m \gamma\mathfrak{E}\left(f^{-1/2}-\frac{v \mathfrak{L}_z}{r \mathfrak{E}}\right). 
\end{equation}
The collision energy becomes:
\begin{equation}
    \mathcal{M}^2=m^2+n^2+2m \gamma\mathfrak{E}\left[(
    1-\frac{e^{-a}}{\rho}
)^{-1/2}-\frac{v l_z}{\rho}\right],
\end{equation}
where $l_z={\mathfrak{L}_z}/{\mathfrak{E} r_h e^a}$. As discussed above, $\gamma_-$ reaches large values. But this case is not interesting in our search. However, in the case $\gamma_+$ has smaller values, but the motion occurs close to the horizon, and the first term $f^{-1/2}$ within parentheses takes on a sufficiently large value. Regarding the second term, we know that
\begin{equation}
    \frac{v l_{z}}{\rho}<l_{z}\leq l_{crit},
\end{equation}
so the second term also has finite small values.
Now, the energy within a center of mass system can reach significant values even with relatively low initial energies for the colliding particles. Under certain conditions, scattering can result in a significant increase in the energy of one particle that significantly exceeds its initial value. This confirms the effectiveness of acceleration mechanisms, although their implementation depends on interaction parameters, such as the energy and direction of particle motion.

\section{Oscillatory Motion of Charged Test Particles Around Schwarzschild-like Black hole}
\label{qpo}

The oscillatory behavior of test particles near gravitationally compact objects is a fundamental aspect of relativistic astrophysics \cite{bardeen1968non, 2016A&A...586A.130S, Remillard:2002cy, 2001AIPC..599..365S, 2014MNRAS.444.2065I}. This motion can be classified into radial and vertical oscillations, both of which provide valuable insights into the dynamics and spacetime structure surrounding these objects. Radial oscillations describe the periodic motion of a particle moving inward and outward along the radial direction, typically occurring around stable circular orbits. The stability and frequency of these oscillations are determined by the spacetime geometry and the properties of the central mass. Vertical oscillations, also called latitudinal or epicyclic oscillations, involve motion perpendicular to the orbital plane. These oscillations provide information about the vertical stability of the orbit and are influenced by the angular momentum and spin of the central object. The vertical epicyclic frequency represents the rate at which a particle oscillates around the equatorial plane of the massive body.

Here, we investigate the oscillatory motion of a massive charged particle near stable circular orbits in the Schwarzschild-like spacetime in the presence of a magnetic field. The four-velocity of the particle is given by:
\begin{align}
  u^\mu = \dot{t} (1, 0, 0, \Omega),
\end{align}
where \(\Omega = d\phi/dt\) represents the particle's angular velocity. 

By using the normalization condition for the four-velocity, the equation of motion takes the form:  
\begin{align}
  g_{rr} \dot{r}^2 + g_{\theta\theta} \dot{\theta}^2 + V(r,\theta) = 0,  
\end{align}
where \(V(r,\theta)\) is the new effective potential.

The radial and vertical oscillation frequencies are given by  
\begin{align}
   \Omega_r^2 = \frac{-g_{tt}-\Omega^2 g_{\phi\phi}}{2g_{rr}} \partial_r^2V\ , \\
   \Omega_\theta^2 = \frac{-g_{tt}-\Omega^2 g_{\phi\phi}}{2g_{\theta\theta}} \partial_\theta^2V\ .
\end{align}

Our results indicate that the vertical frequency \(\Omega_\theta\) coincides with the orbital angular velocity \(\Omega\). All fundamental frequencies are expressed in Hertz by incorporating fundamental constants:  
\begin{align}
   \nu_i = \frac{1}{2\pi} \frac{c^3}{G M} \Omega_i, 
   \label{frq}
\end{align}
where \(c = 3 \times 10^{10} \, \text{cm s}^{-1}\) is the speed of light, and \(G = 6.67 \times 10^{-8} \, \text{cm}^3 \text{g}^{-1} \text{s}^{-2}\) is the gravitational constant.  

\begin{figure*}
    \centering
    \includegraphics[width=0.45\linewidth]{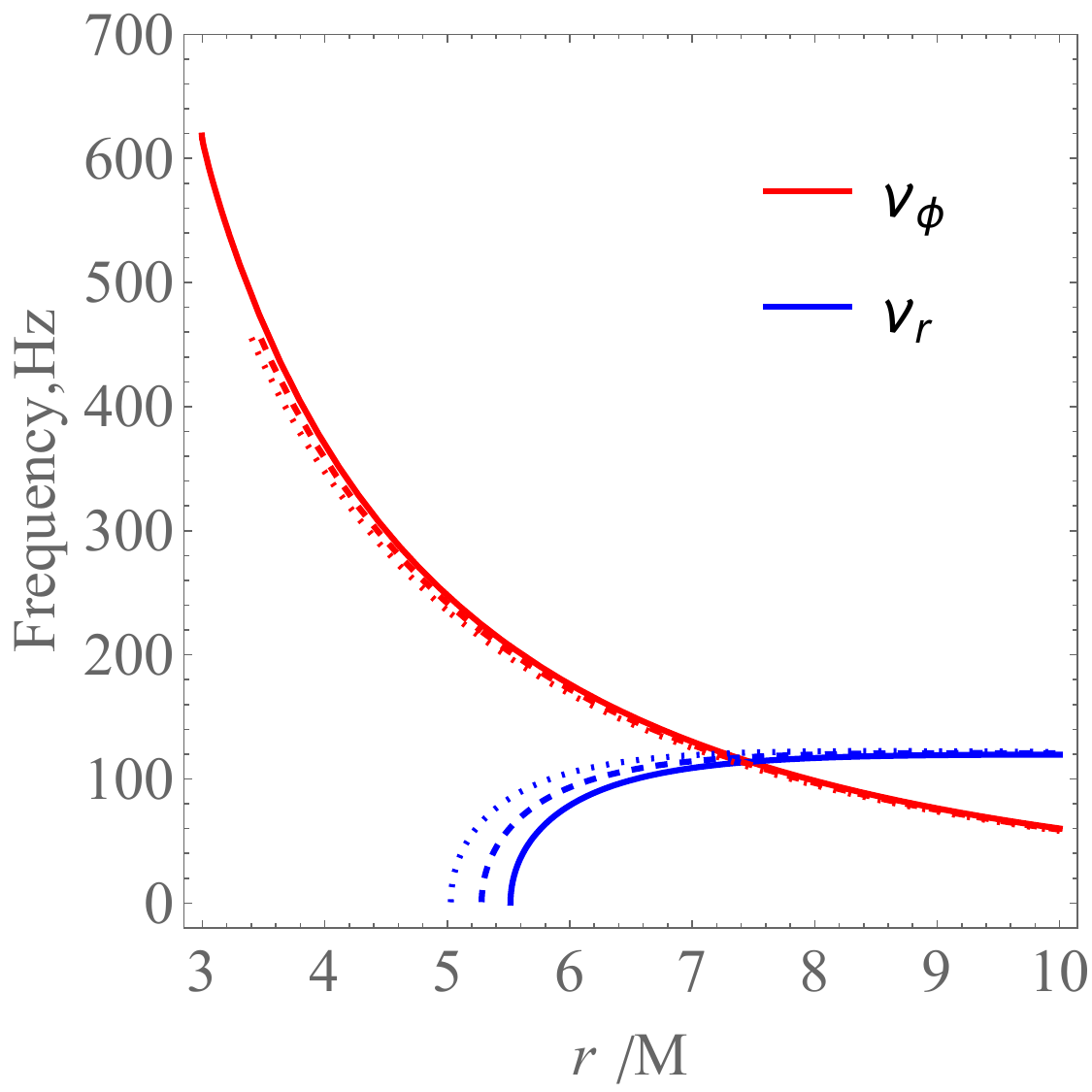}
    \includegraphics[width=0.45\linewidth]{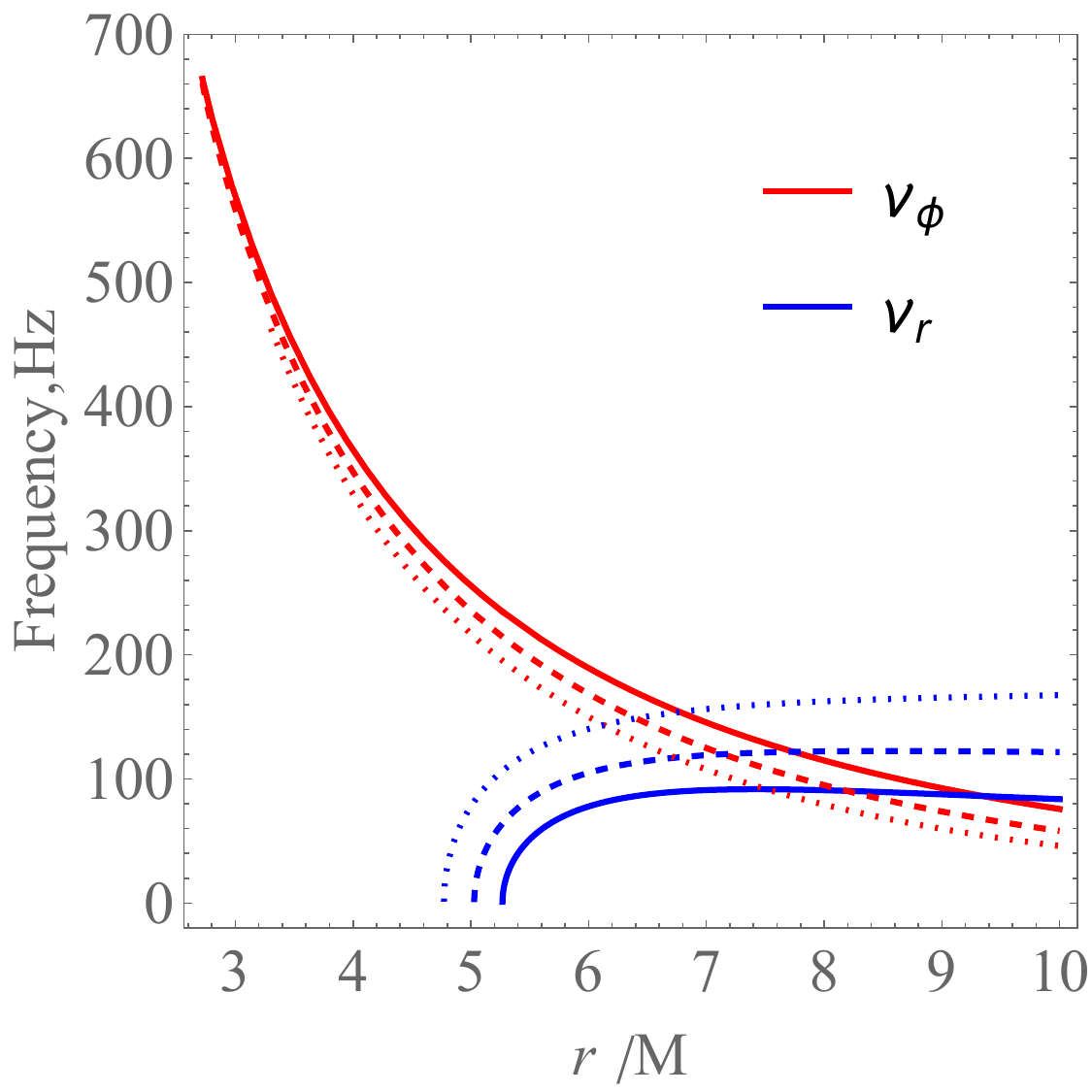}
    \caption{\justifying Radial dependency of Fundamental frequencies of massive charged particle in external magnetic field: Left graph for fixed magnetic value, $0.02$ (Solid line $a/M=0.0$, dashed line $a/M=0.1$, dotted line $a/M=0.2$). Right panel for fixed spacetime paremater $a=0.2$ (Solid line $b=0.01$, dashed line $b/M=0.02$, dotted line $b/M=0.03$). }
    \label{sfr}
\end{figure*}

The radial dependence of the fundamental frequency of massive charged particles in the external magnetic field represent in Fig.~{\ref{sfr}}. The left panel of the graph illustrates that as the metric parameter $a$ increases, so do the frequencies. Near the massive object, the difference between the radial and vertical components of the frequency is significant. However, as the distance from the mass increases, this difference decreases. In the large region $r$, the frequency lines merge, indicating a weakening of the gravitational and magnetic effects. In contrast, the right panel shows that as the magnetic parameter $b$ increases, the radial and vertical frequencies begin to diverge. This divergence is due to the increased strength of the magnetic interaction, which affects the particle dynamics.

\subsection{Quasi-Periodic Oscillations}

The interplay between orbital, radial, and vertical oscillations results in intricate trajectories for testing charged particles. These oscillations play a key role in understanding QPOs, which are commonly observed in X-ray binaries. QPOs are thought to arise from the oscillatory motion of matter within the accretion disk surrounding a black hole or neutron star. This motion influences both the stability and structure of the accretion disk. By analyzing these oscillations, researchers can model the emission spectra and variability of accretion disks. Furthermore, studying the frequencies and modes of oscillations provides valuable information about the properties of black holes and neutron stars, including their mass, spin, and geometry of the surrounding spacetime.  

Since astrophysical compact objects such as black holes do not emit electromagnetic radiation directly, their presence is inferred through their gravitational influence. The intense gravitational field around a black hole distorts spacetime, significantly affecting the motion of nearby matter. This effect leads to the formation of an accretion disk, a rotating mass of gas, dust, and other material captured by the gravitational pull of the black hole. As matter within the disk spirals inward, friction and other forces generate extreme heat and pressure, causing the disk to emit radiation across a wide range of wavelengths. Although black holes themselves remain invisible, the luminous emissions from their accretion disks provide crucial insights into their existence and properties.  

The RP model serves as a theoretical framework to explain the origin of QPOs. According to this model, QPOs arise from the quasi-harmonic oscillations of charged particles as they undergo radial and angular motion around black holes and wormholes. The RP model establishes a direct relationship between the dynamics of these particles and the emergence of QPO phenomena, offering a deeper understanding of the oscillatory mechanisms in strong gravitational fields.  

In the RP model, twin-peaked QPOs are interpreted as follows: the higher frequency component corresponds to the orbital frequency of the particle, denoted as \( \nu_U = \nu_\phi \), while the lower frequency component is given by the difference between the orbital frequency and the radial oscillation frequency, expressed as \( \nu_L = \nu_\phi - \nu_r \). In essence, the upper frequency represents the orbital motion of the particle, while the lower frequency arises from the interaction between the orbital and radial oscillations. This interpretation provides a clear framework for analyzing the frequency components observed in twin-peaked QPOs, offering valuable insights into the dynamics of astrophysical systems in strong gravitational fields.

\begin{figure*}
    \centering
    \includegraphics[width=0.45\linewidth]{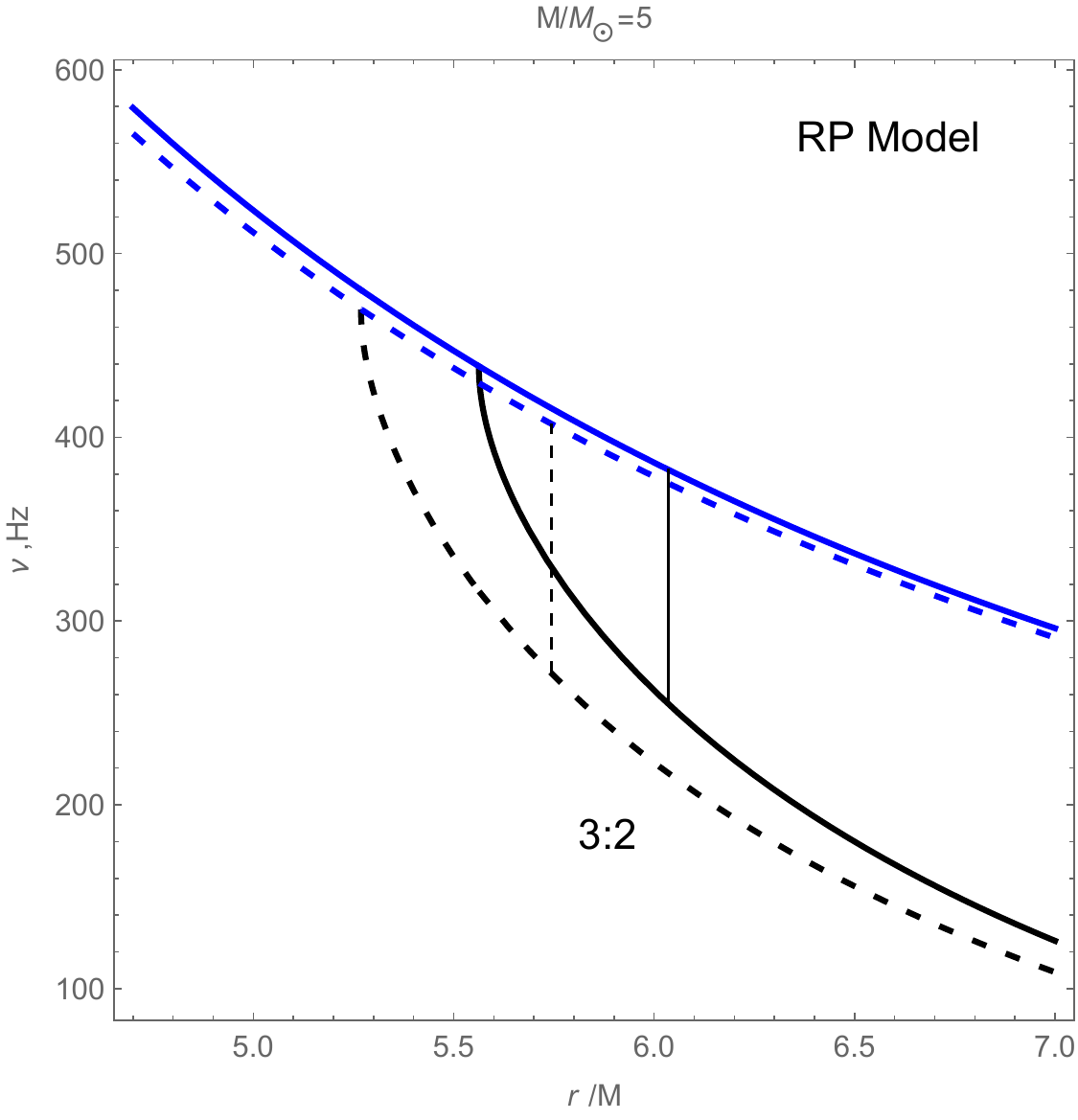}
    \includegraphics[width=0.45\linewidth]{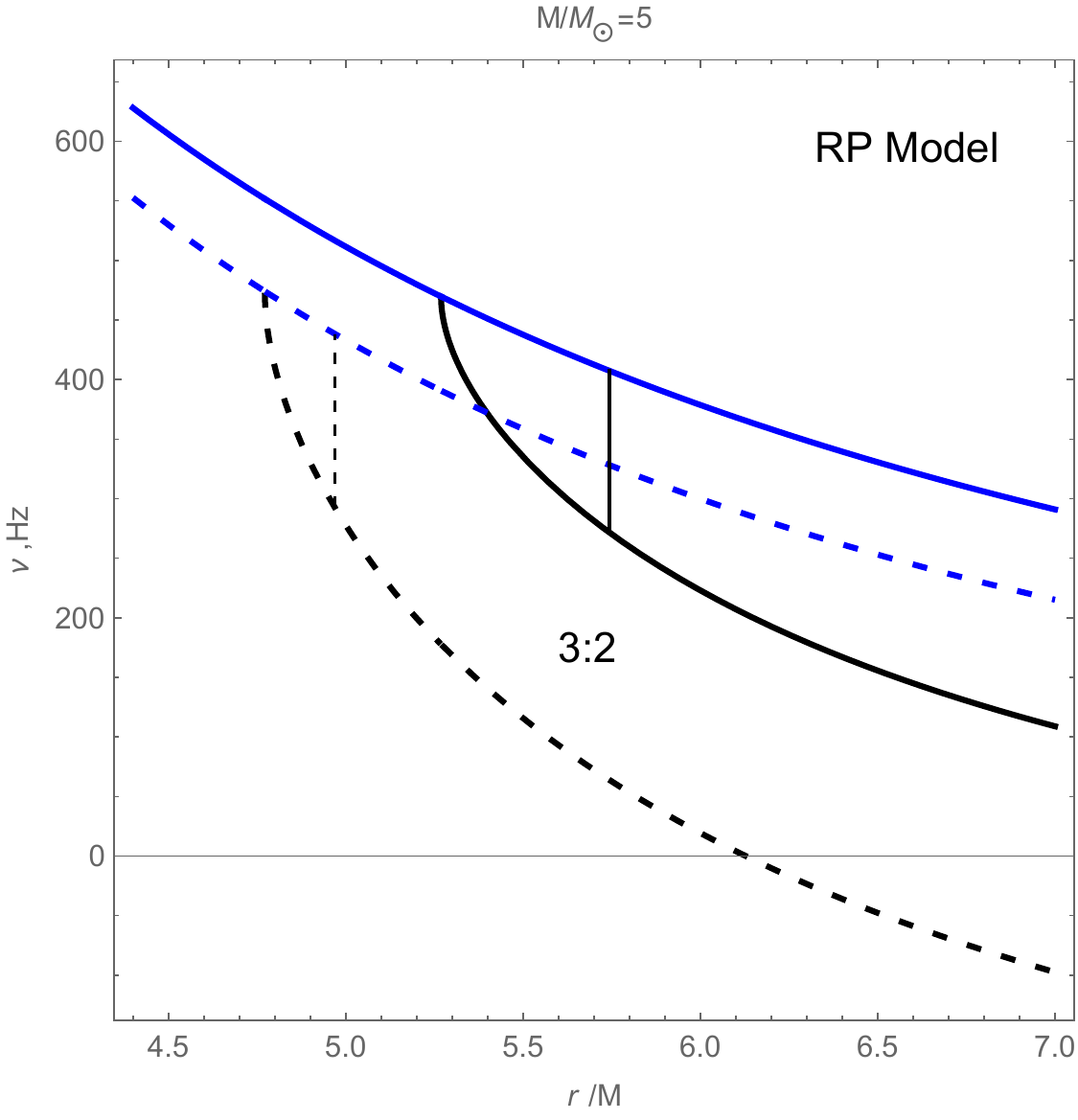}
    \caption{\justifying Upper (blue lines) and Lower (black lines) frequencies with $\nu_U/\nu_L=3/2$ ratio. Left graph for fixed $b/M=0.01$, solid lines $a/M=0.1$ and dashed lines $a/M=0.2$, respectively. Right graph shows for fixed $a/M=0.2$ and different values of  magnetic field ($b/M=0.01$ and $b/M=0.03$) are given solid and dashed lines.}
    \label{s32}
\end{figure*}

\begin{figure*}
    \centering
    \includegraphics[width=0.45\linewidth]{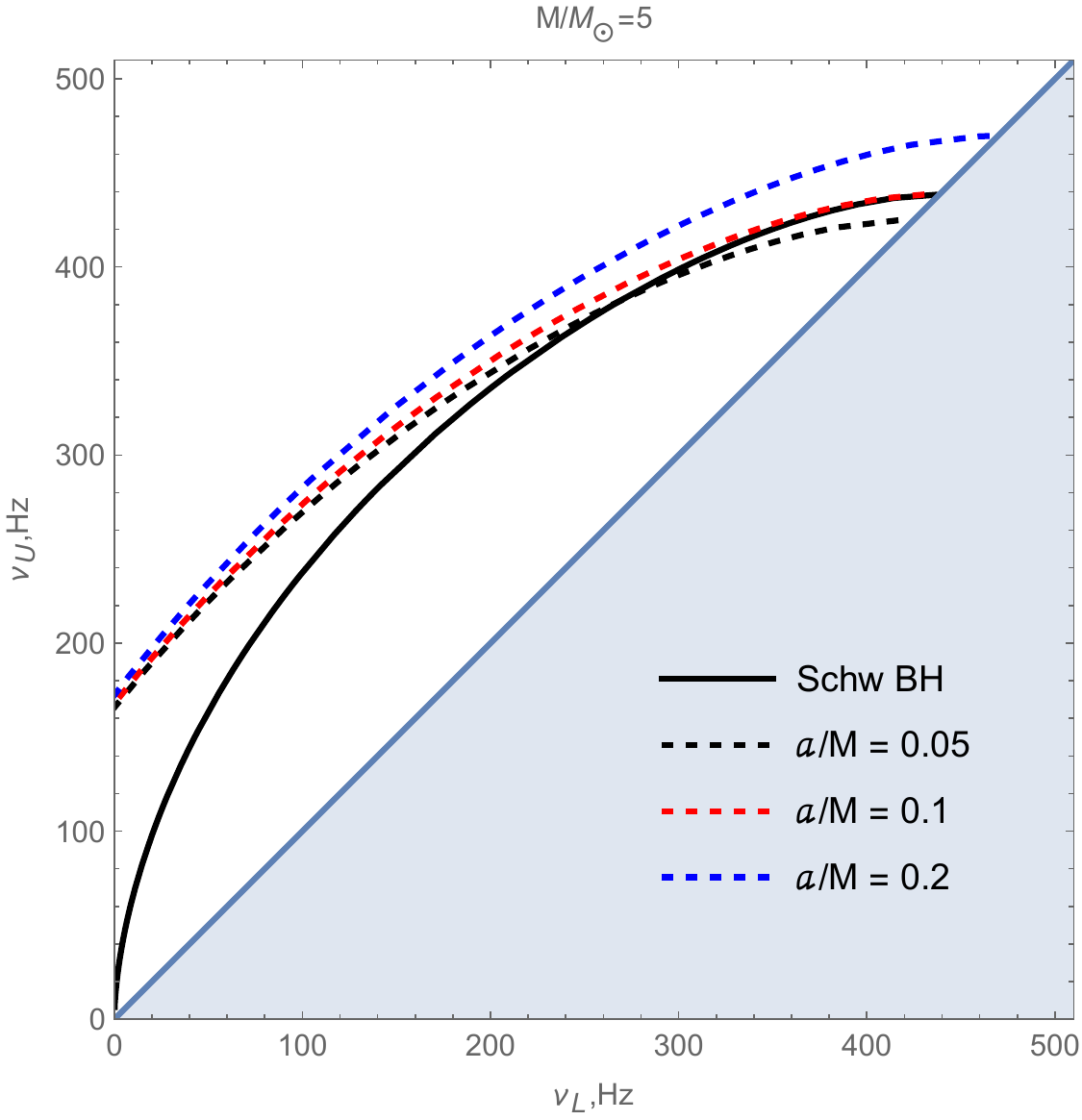}
    \includegraphics[width=0.45\linewidth]{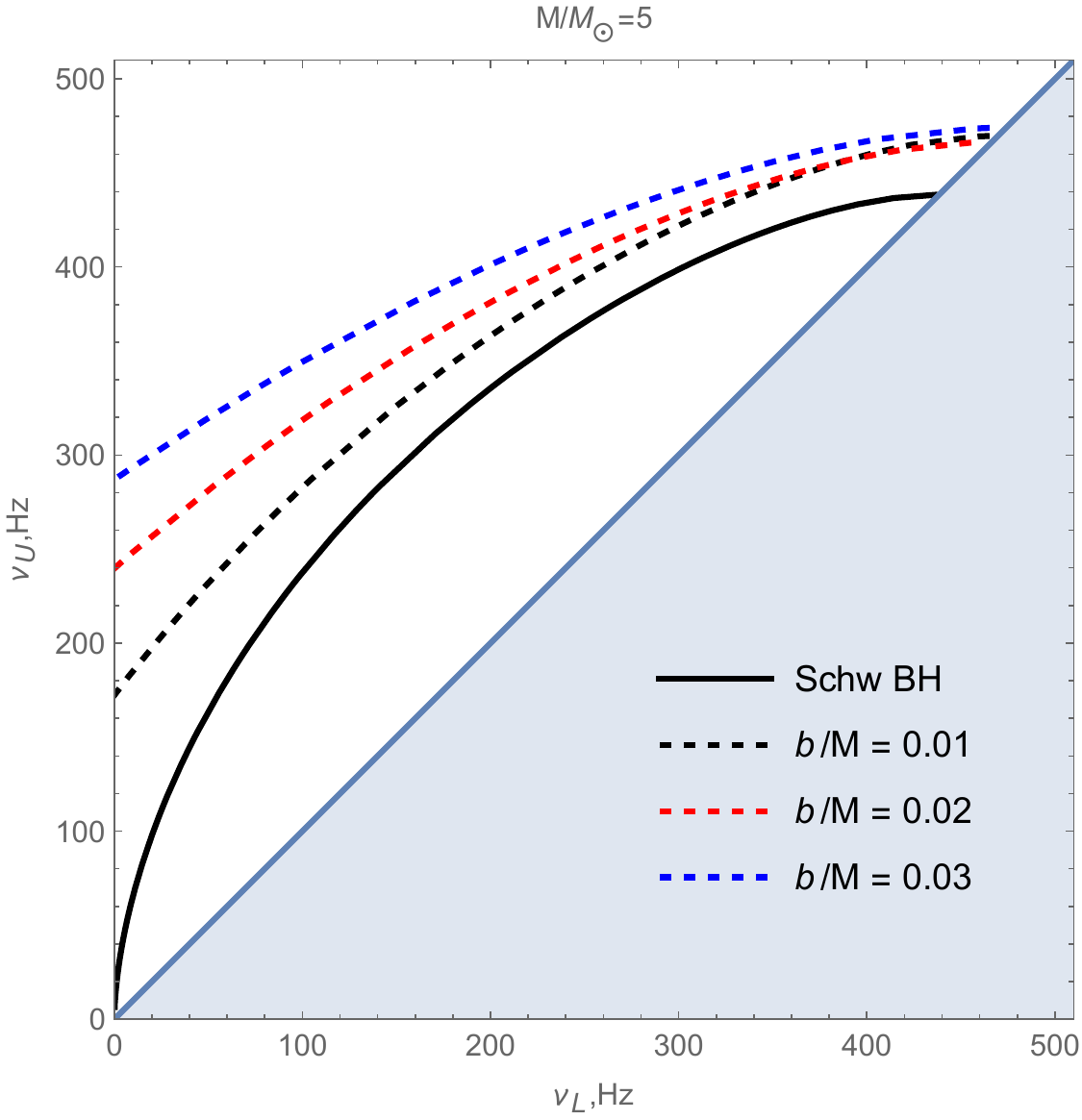}
    \caption{\justifying Relationship between the upper $\nu_U$ and lower $\nu_L$ frequencies of twin-peak QPOs in the RP model. Left panel for different values of the metric parameter $a$ for a fixed value $b/M=0.01$ and right panel for different values of the metric parameter $b$ for a fixed value $a/M=0.2$.}
    \label{rpmodel}
\end{figure*}

Figure~{\ref{s32}} presents the upper $\nu_U$ and lower $\nu_L$ QPO frequencies within the RP model, satisfying the resonance condition $\nu_U/\nu_L = 3/2$. The left panel illustrates the variation of frequencies for a fixed magnetic parameter, $b/M = 0.01$, while considering different values of the metric parameter, $a/M = 0.1$ and $a/M = 0.2$. As the metric parameter increases, the upper frequency remains nearly unchanged, whereas the lower frequency shifts significantly towards smaller radial values. The right panel demonstrates the effect of varying the magnetic parameter on the QPOs frequencies while keeping the metric parameter fixed at $a/M = 0.2$. A stronger magnetic field, $b/M = 0.03$, increases the difference between the upper and lower frequencies, shifting the resonance point. The intersection point of the black and blue curves indicates the radius at which the resonance condition is satisfied. 

Figure~{\ref{rpmodel}} depicts the $\nu_U - \nu_L$ diagram for twin-peak QPOs around a black hole within the RP model. The figure is generated for a black hole of stellar mass with \(M = 5M_{\odot}\), and the frequency values in hertz are obtained using Eq.~(\ref{frq}). The shaded region in both panels of Fig.~{\ref{rpmodel}} represents the range of physically meaningful QPO frequencies, where such QPOs cannot be observed. The inclined boundary line of these regions corresponds to the condition in which the upper and lower frequencies become equal, causing the two peaks in the QPO spectrum to merge into a single peak. This implies that if a QPO position falls below this boundary within the shaded region, the QPO phenomenon becomes unobservable. The right panel illustrates this dependence for a fixed magnetic parameter $b$ while varying the metric parameter $a$. As $a$ increases, the deviation from the Schwarzschild case becomes more pronounced. Notably, for small oscillations, the differences between the curves are minor, but they grow as the oscillation amplitudes increase. The left panel, on the other hand, shows the dependence for a fixed metric parameter $a$ while varying the magnetic parameter $b$. Here, the opposite trend is observed: for small oscillations, the differences between the curves are significant, but they converge for larger oscillations. This can be explained by the fact that the magnetic parameter $b$ compensates for the difference. As seen in Fig.~{\ref{fig:multiple_graphs}}, at larger distances from the massive object, the magnetic field becomes uniform, which accounts for the merging of lines at higher frequencies, as the interaction between the charged particle and the massive object weakens.

\section{Conclusion}\label{Sec:conclusions}

In this paper, we have focused on examining the geodesic motion within a Schwarzschild-like spacetime, also known as the ``eye of the storm", with a particular focus on the influence of the parameter $a$ obtained in \cite{Culetu:2014lca, Culetu:2015cna}. We have found analytical expressions for the circular motion of massive particles within this spacetime. We have also explored the ISCO around the black hole. The motion of charged particles near the Schwarzschild-like black holes has been studied in the presence of an external, asymptotically homogeneous magnetic field using the Hamilton-Jacobi equations. The study of electromagnetic fields has shown that the angular component of the magnetic field, i.e., the radial component, increases with increasing parameters \(a\). We have also illustrated the configuration of magnetic field lines depending on the magnitude of the involved parameter and compared the values of circular orbits for charged and neutral particles. The motion of the particles has been found to vary significantly depending on changes in the value of the magnetic parameter $b \ll 1$. 

In the current analysis,  we have focused on the limited trajectories of charged particles. Such trajectories can be considered as an approximation of the motion of charged particles in the accretion disk of a black hole when their mutual interaction is negligible. We have constructed an approximate solution to the dynamic equations: a limited trajectory localized near a stable circular orbit.
It is also worth noting that the additional parameter $a$ in the Schwarzschild-like spacetime metric \cite{Simpson:2019mud,Simpson:2021dyo} simulates a decrease in mass of a black hole, and this effect can be observed in our research. In the case of orbital motion, this proves that the parameter helps the particle remain in a stable orbit with a closer radius. As a result, a particle in a stable orbit at a closer radius requires less energy and angular momentum. The effect of this parameter also applies to the particle speed, allowing it to move steadily in a radius that is closer to it. The larger the parameter $a$, the more noticeable its effect.

We analyze the properties of the corresponding effective potential arising from the combined effect of gravitational and Lorentz forces on a charged particle in this metric by using the method described in Ref.~\cite{Frolov11}.  We have found that the metric parameter $a$ has an effect that reduces all the data that can be achieved for the Schwarzschild black hole. The potential for the Schwarzschild-like black hole metric also has two extreme points, so in this critical case, we have determined the position of the inner boundary of a stable circular orbit. 

We have alos found that the collision energy is not high when two particles collide along the same trajectory near the ISCO in opposite directions. However, it can be high if a particle falling from infinity collides with a charged particle orbiting around the ISCO.  A black hole can be an effective particle accelerator, contributing to large values of collision energy, but the implementation of this process depends on the interaction parameters (energy and direction of motion of the colliding particles). Our analysis is somewhat simplified since we have used a simple particle motion and collision model without considering plasma and other related effects.

We have also studied QPOs around Schwarzschild-like black holes as an application of harmonic oscillations in the RP model, along with the possible values of the upper and lower frequencies of twin-peak QPOs and the radius of the QPO orbit at a 3:2 frequency ratio. We have analyzed the dependence of the oscillations on the magnetic parameter $b$ and the metric parameter $a$, describing how each of them affects the dynamics of the test charged particle. It has been found that QPOs orbits and the ISCO are close to each other in the RP model. This suggests that the ISCO measurement problem can be addressed through the study of twin-peak QPOs within the framework of the RP model.

Overall, this study highlights the influence of the additional parameter $a$ on particle dynamics in Schwarzschild-like black holes. Future research should explore the implications of this parameter for other astrophysical processes, including plasma effects and accretion disk dynamics.

\section*{Acknowledgments}
This research was supported by the Grants F-FA-2021-510 from the Uzbekistan Ministry for Innovative Development.

\bibliographystyle{apsrev4-2}  
\bibliography{Ref,asddd,asddd1}
\end{document}